\documentclass[letterpaper, 10 pt, conference]{ieeeconf}

\IEEEoverridecommandlockouts
\usepackage{graphicx}
\usepackage[T1]{fontenc}
\usepackage{amsmath}
\usepackage{amssymb}
\usepackage{threeparttable}
\makeatletter
\let\NAT@parse\undefined
\makeatother
\usepackage{hyperref}

\title{\LARGE \bf
BSS-CFFMA: Cross-Domain Feature Fusion and Multi-Attention Speech Enhancement Network based on Self-Supervised Embedding*
}

\author{Alimjan Mattursun$^{1}$, Liejun Wang$^{1\dagger}$\thanks{$^{\dagger}$ Both Liejun Wang and Yinfeng Yu are corresponding authors.} and Yinfeng Yu$^{1\dagger}$% <-this % stops a space
\thanks{*This work was supported by these works: the Tianshan Excellence Program Project of Xinjiang Uygur Autonomous Region, China (2022TSYCLJ0036); the Central Government Guides Local Science and Technology Development Fund Projects (ZYYD2022C19); the National Natural Science Foundation of China under Grant 62303259.}
\thanks{$^{1}$Alimjan Mattursun, Liejun Wang, and Yinfeng Yu are with the School of Computer Science and Technology, Xinjiang University, Urumqi 830049, China (e-mail:{\tt\small alim@stu.xju.edu.cn;
        wljxju@xju.edu.cn;
        yuyinfeng@xju.edu.cn;).}}
}

\begin{document}
\maketitle
\thispagestyle{empty}
\pagestyle{empty}

\begin{abstract}
Speech self-supervised learning (SSL) represents has achieved state-of-the-art (SOTA) performance in multiple downstream tasks. However, its application in speech enhancement (SE) tasks remains immature, offering opportunities for improvement. In this study, we introduce a novel cross-domain feature fusion and multi-attention speech enhancement network, termed BSS-CFFMA, which leverages self-supervised embeddings. BSS-CFFMA comprises a multi-scale cross-domain feature fusion (MSCFF) block and a residual hybrid multi-attention (RHMA) block. The MSCFF block effectively integrates cross-domain features, facilitating the extraction of rich acoustic information. The RHMA block, serving as the primary enhancement module, utilizes three distinct attention modules to capture diverse attention representations and estimate high-quality speech signals.

We evaluate the performance of the BSS-CFFMA model through comparative and ablation studies on the VoiceBank-DEMAND dataset, achieving SOTA results. Furthermore, we select three types of data from the WHAMR! dataset, a collection specifically designed for speech enhancement tasks, to assess the capabilities of BSS-CFFMA in tasks such as denoising only, dereverberation only, and simultaneous denoising and dereverberation. This study marks the first attempt to explore the effectiveness of self-supervised embedding-based speech enhancement methods in complex tasks encompassing dereverberation and simultaneous denoising and dereverberation. The demo implementation of BSS-CFFMA is available online\footnote[2]{https://github.com/AlimMat/BSS-CFFMA. \label{s1}}.
\end{abstract}

\section{INTRODUCTION}
In everyday acoustic environments, various forms of background noise and room reverberation significantly degrade the clarity and intelligibility of speech, posing significant challenges for speech-related applications such as conferencing systems, speech recognition systems, and speaker recognition systems \cite{review1}. Speech enhancement (SE) tasks aim to extract clean speech from noisy speech and improve the quality and intelligibility of speech. Recently, researchers have investigated deep neural network (DNN) models for speech enhancement. DNN models have shown powerful denoising capabilities in complex noise environments compared to traditional methods  \cite{review2}.

With the development of DNN, significant progress has been made in single-channel speech enhancement tasks. DNN-based SE methods can be broadly categorized into time-domain approaches \cite{SEGAN,WAVCRN,TSTNN,MANNER,SADN-UNet}, time-frequency (T-F) domain approaches \cite{METRICGAN,METRICGAN+,DMFNET,COMPNET}, and cross-domain approaches \cite{SSL2,BSSSE,SSL6}. Time-domain methods directly estimate the target clean speech waveform from the noisy speech waveform. Time-frequency domain methods estimate clean speech from the spectrogram generated by applying the short-time Fourier transform (STFT) to the original signal. Cross-domain methods process features from various speech domains to capture more acoustic information about speech and noise, facilitating the estimation of clean speech \cite{SSL2,BSSSE}.

Self-supervised learning (SSL) leverages many unlabeled data to extract meaningful representations  \cite{review3}. In many applications, supervised learning is generally superior to unsupervised learning. However, collecting a large amount of labeled data is time-consuming and sometimes impractical. SSL has been validated in various domains and has improved the performance of downstream tasks. Specifically, some promising SSL models have been proposed for speech-related tasks, such as speech and emotion recognition. As of now, there are many speech SSL models available, with the best-performing ones including
Wav2vec2.0 \cite{WAV2VEC20}, WavLM \cite{WAVLM}, HuBERT \cite{HUBERT} and others.
However, there is relatively little research on the application of SSL features to SE. Huang et al. \cite{SSL3} proposed the application of SSL features to SE and comprehensively evaluated the performance of most SSL models in SE. Hung et al. \cite{BSSSE} employed a weight-summed SSL framework, fusing SSL features with spectrograms to address the issue of fine-grained information loss in SSL features. However, their cross-domain feature fusion method using early concatenation (concat) may limit the enhancement performance due to insufficient cross-domain feature integration \cite{EARLYCONCET}. In addition, previous studies on self-supervised embedding-based methods for speech enhancement \cite{SSL2, BSSSE, SSL3} commonly employed simple RNN-based models for the enhancement module, while recent attention-based enhancement architectures \cite{TSTNN, MANNER, UFORMER} have demonstrated strong denoising capabilities in speech enhancement.

In this paper, we propose a cross-domain feature fusion and multi-attention speech enhancement network based on self-supervised embedding (BSS-CFFMA). We design a multi-scale cross-domain feature fusion module (MSCFF) in BSS-CFFMA to better fuse self-supervised features and spectrogram features, extracting information at different granularities, and further addressing the issues of SSL information loss and insufficient feature fusion \cite{BSSSE,jiao2024mfhca}. Additionally, we design a residual-mixed multi-attention module (RHMA) in BSS-CFFMA, which incorporates a selective channel-time attention fusion module (SCTA) using a self-attention design to obtain different attention feature representations and achieve improved speech enhancement.
\begin{figure*}[htbp]
  \centering
  \includegraphics[width=18cm]{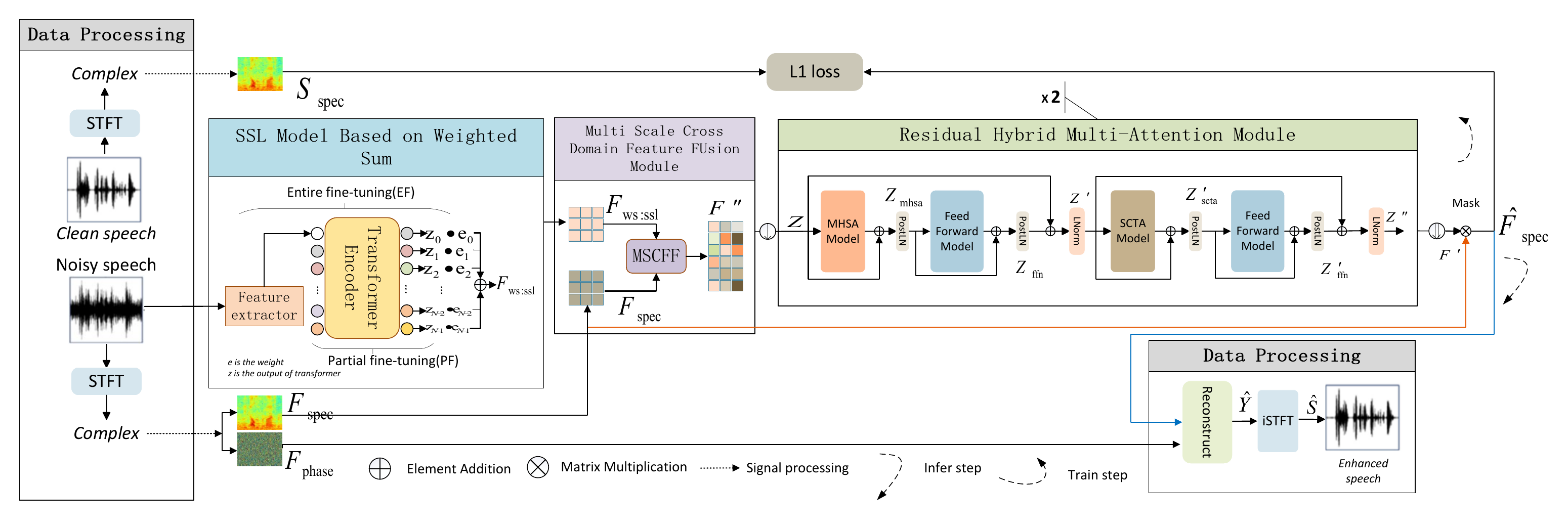}
  \caption{Architecture of the proposed BSS-CFFMA. "STFT" represents the short-time Fourier transform of the speech. "Spec" stands for Spectrogram. \textcircled{$\downarrow$} represents the downsampling operation which Liner and Z $\in \mathbb{R}^{T*512}$. \textcircled{$\Downarrow$} represents the downsampling operation which Liner and Sigmod. "iSTFT" represents the inverse short-time Fourier transform of the speech. "Reconstruct" represents the reconstruction of the speech complex spectrum for the speech spectrogram $\in \mathbb{R}^{F*T}$  and phase $\in \mathbb{R}^{F*T*2}$.}
  \label{fig:f1}
\end{figure*}

\section{RELATED WORK}
\subsection{SSL Model}
The SSL models can be categorized into generative modeling, discriminative modeling, and multi-task learning. Generative modeling reconstructs input data using an encoder-decoder structure. Multi-task learning involves learning multiple tasks simultaneously, where the model can extract features that are useful for all the tasks through shared representations. Discriminative modeling maps input data to a representation space and measures the corresponding similarity. 
In this study, we utilized two base SSL models to extract latent representations: Wav2vec2.0 (Base) and WavLM (Base).

\subsection{Cross Domain Features and Fine Tuning SSL}
Studies \cite{SSL6} and \cite{BSSSE} have shown that cross-domain features contribute to improving the performance of automatic speech recognition (ASR) and speech enhancement (SE). 
Studies \cite{SSL3} have shown that SSL has great potential in speech enhancement tasks. However, Studies \cite{BSSSE} adopted weighted sum SSL and fine-tuning methods, significantly improving the performance of speech enhancement. In this study, we employ SSL and Speech Spectrogram as two cross-domain features, weighted summed SSL and a more efficient partially fine-tuned (PF) approach to improve the performance of speech enhancement further.

\section{METHOD}
Fig. \ref{fig:f1} illustrates the overall architecture of BSS-CFFMA, which consists of an SSL model with weighted summation, a multi-scale cross-domain feature fusion (MSCFF) module, and two residual hybrid multi-attention (RHMA) modules.

Firstly, noisy speech is fed into a weighted sum SSL model and STFT to generate SSL latent representations $F_{ws:ssl}$ and spectrograms $F_{spec}$, respectively. Subsequently, the $F_{ws:ssl}$ and $F_{spec}$ features are input into the MSCFF module for feature fusion across domains, resulting in the feature $F^{''}$. $F^{''}$ is then fed into RHMA, yielding different attentional representations through various attention mechanisms. Ultimately, the enhanced spectrogram is obtained by element-wise multiplication of the output from the second RHMA with the noisy spectrogram. During inference, the enhanced spectrogram and noise phase are utilized to reconstruct the enhanced speech waveform.

\subsection{SSL Model based on Weighted Sum}
In study \cite{BSSSE}, the author believes that using the last layer of SSL directly may result in the loss of some local information necessary for speech reconstruction tasks in deeper layers. So learnable parameter $ e(i) $ is designed for each transformer layer's output $ z(i) $ in SSL:
\begin{equation}
F_{ws:ssl} = \sum_{i=0}^{N-1} \left[ e(i)*z(i) \right],
\end{equation}
where $F_{ws:ssl}\in \mathbb{R} ^{D*T}$, i=0$ \cdots $N-1 is the number of layers in SSL. Parameters $0 \le e(i) \le 1, \sum_{i} e(i) = 1 $.

\subsection{Multi Scale Cross Domain Feature Fusion (MSCFF)}
In study \cite{BSSSE}, complemented fine-grained information by incorporating the original acoustic features on top of SSL, resulting in improved performance.
It uses the early concatenation (Concat). In contrast, \cite{EARLYCONCET} shows that early Concat focuses the entire cross-modal fusion process on a single modality and reduces feature diversity and fine-grained information. However, multi-scale feature extraction and fusion strategies have been shown to efficiently integrate cross-modal features, significantly enhancing network performance \cite{yu2021weavenet}.
Considering the research findings and aiming to better integrate and extract information from SSL and spectrogram features, we introduce the multi-scale cross-domain feature fusion (MSCFF) module.

\begin{figure}[htbp]
  \centering
  \includegraphics[width=9cm]{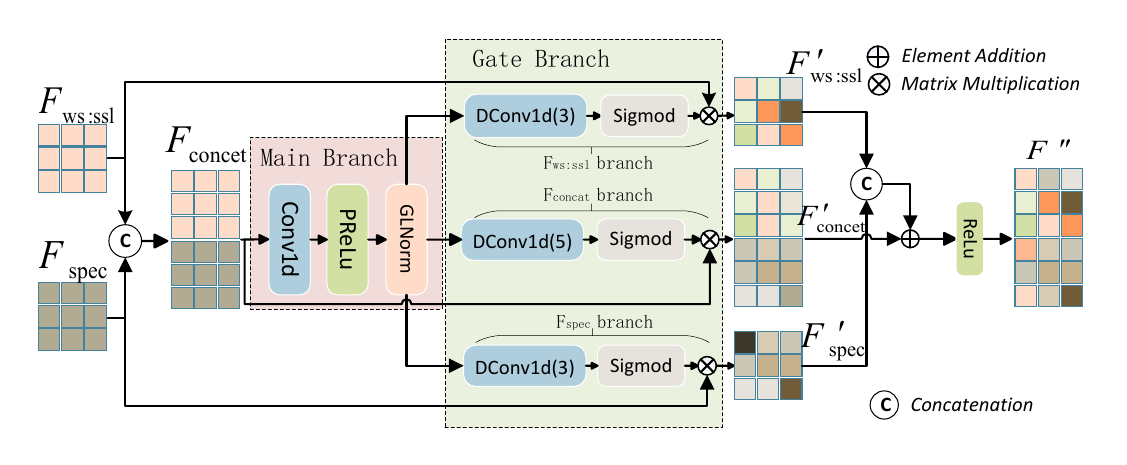}
  \caption{ Structure multi-scale cross-domain feature fusion (MSCFF) model. "DConv1d" represents the dilation convolution, where the kernel sizes are 3 and 5, respectively, and the dilation coefficients are 1.}
  \label{fig:f2}
\end{figure}
The architecture of the MSCFF model is illustrated in Fig. 2, comprising a main branch (MB) and three gate branches (GB). The main branch, along with one gate branch, forms a classic STCM \cite{STCM} structure. The main branch consists of a 1D convolutional layer, a Prelu activation function, and layer normalization (LNorm). The gate branches comprise dilated convolutional kernels with different sizes and sigmoid activation functions.

The process begins by concatenating the SSL feature $F_{ws:ssl}$ and the feature $F_{spec}$ to obtain the fused feature F$_{concet}$. The $F_{concet}$ is then fed into the main branch for feature extraction, resulting in the output $F^{'}$.
\begin{equation}
F^{'}=MB(concat(F_{ws:ssl},F_{spec})),
\end{equation}
subsequently, $F^{'}$ is passed through the gate branch.
\begin{equation}
F^{'}_{spec,concet,ws:ssl} = GB(F^{'})*F_{spec,concat,ws:ssl},
\end{equation}
finally, the three features are cross-fused.
\begin{equation}
F^{''}=ReLu(concat(F^{'}_{spec},F^{'}_{ws:ssl})+F^{'}_{concet}),
\end{equation}
where 
$F_{spec} \sim F^{'}_{spec} \in \mathbb{R}^{F * T}$,
$F_{ws:ssl} \sim F^{'}_{ws:ssl} \in \mathbb{R}^{D * T}$, $F_{concet} \sim F^{'}_{concet} \sim F^{''} \in \mathbb{R}^{(D+F) * T}$.

\subsection{Residual Hybrid Multi-Attention (RHMA) Model}
In previous studies \cite{SSL2,SSL3,BSSSE,SSL5}, RNNs were commonly used as the primary speech enhancement module for self-supervised embedding. However, RNNs suffer from long-term dependency issues, high parameter counts, and low computational efficiency. Recently, models based on Transformer architecture have achieved remarkable performance in the field of speech recognition, such as 
Squeezeformer \cite{SQUEEZEFORMER}, among others. In the domain of speech enhancement, utilizing self-attention modules often leads to improved performance, as observed in TSTNN \cite{TSTNN}, Uformer \cite{UFORMER}, and similar works.
Based on the aforementioned research, in order to obtain more useful information and enhance performance in cross-domain feature fusion, we designed a residual hybrid multi-attention (RHMA) module.

The structure of the RHMA module is shown in Fig. \ref{fig:f1}. It is based on the architecture of Squeezeforme \cite{SQUEEZEFORMER}. The module consists of a multiple-head self-attention (MHSA) block, a feed-forward (FFN) block, and a selective channel-time attention (SCTA) fusion block. Post-layer normalization (PostLN) is employed between the blocks for normalization, and multiple residual connections are utilized to optimize the model structure, facilitating rapid convergence.

The fused feature Z is obtained after the MSCFF module is fed into the MHSA module.
\begin{equation}
Z_{mhsa}=PostLN(MHSA(Z)+Z),
\end{equation}
$Z_{mhsa}$ represents the output of the MHSA block, which is passed through a residual connection and PostLN. 
\begin{equation}
Z^{'}=LN(PostLN(FFN(Z_{mhsa})+Z_{mhsa})+Z),
\end{equation}
$Z^{'}$ represents the output of the FFN, which undergoes two levels of residual connections and Layer Normalization.
\begin{equation}
Z^{'}_{scta}=PostLN(SCTA(Z^{'})+Z^{'}),
\end{equation}
$Z^{'}_{scta}$ represents the output of the SCTA block, which is passed through a residual connection and PostLN.
\begin{equation}
Z^{''}=LN(PostLN(FFN(Z^{'}_{scta})+Z^{'}_{scta})+Z^{'}),
\end{equation}
$Z^{''}$ represents the output of the FFN, which undergoes two levels of residual connections and Layer Normalization.

\subsection{Selective Channel-Time Attention Fusion (SCTA) Module}
While models that combine attention and convolution, such as
Squeezeformer \cite{SQUEEZEFORMER}, have achieved remarkable performance in various speech tasks, the convolutional modules increase the parameter count. Research suggests that multi-perspective attention outperforms single attention. Convolutional block attention module \cite{CBAM} (CBAM) is a lightweight and efficient convolutional attention method. In the domain of speech enhancement, CBAM has been utilized as a residual block \cite{HOMOMORPHIC}, yielding excellent performance\cite{guo2023rethinking}. Based on the aforementioned research findings, we have designed the selective channel-time attention (SCTA) fusion module, which has a lower parameter count while capturing information dependencies along the channel and time axes, leading to higher performance.
\begin{figure}[htbp]
\begin{minipage}[b]{1.0\linewidth}
  \centering
  \centerline{\includegraphics[width=9cm]{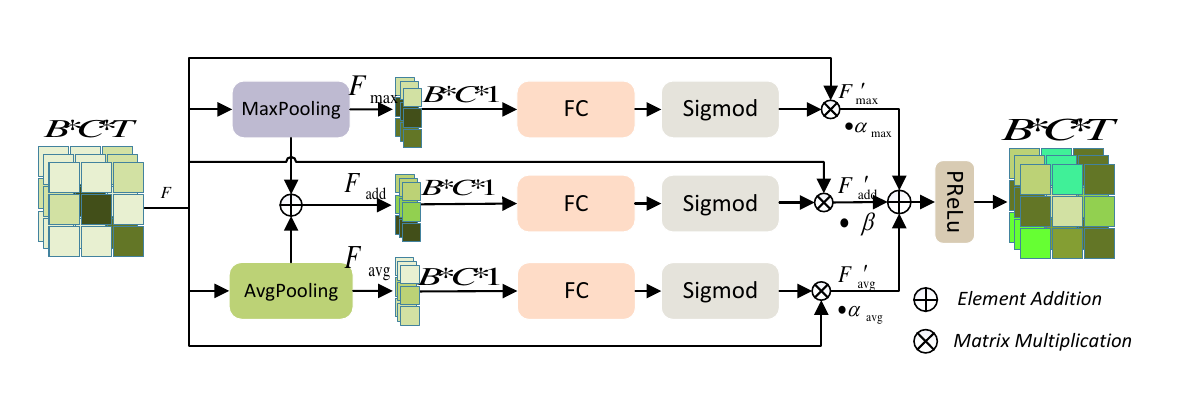}}
\end{minipage}
\caption{Structure of selective channel-attention fusion (SCA) block. Where FC consists of two Liner and one Relu activation function.}
\label{fig:f3}
\end{figure}

\begin{figure}[htbp]
\begin{minipage}[b]{1.0\linewidth}
  \centering
  \centerline{\includegraphics[width=9cm]{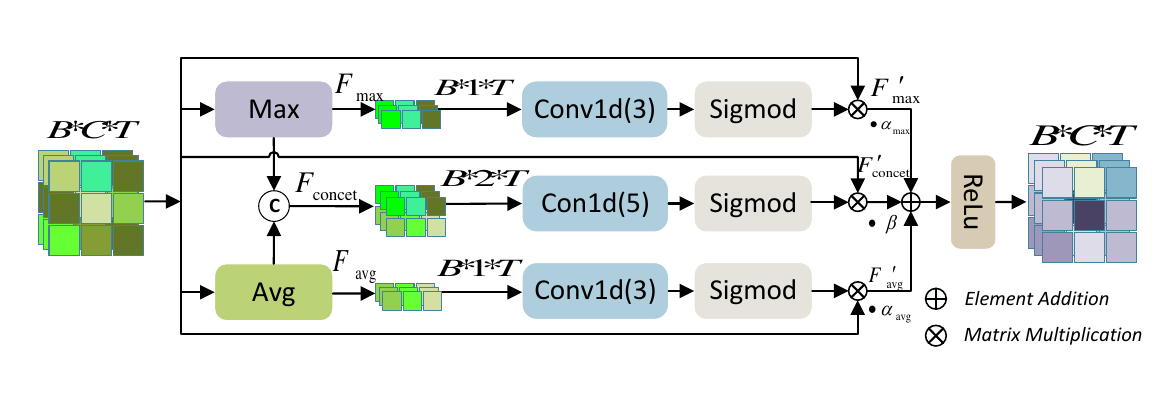}}
\end{minipage}
\caption{Structure of selective time-attention fusion (STA) block. Where the Convolution kernel sizes are 3 and 5, respectively.}
\label{fig:f4}
\end{figure}

The SCTA module consists of two components: selective channel-attention fusion (SCA) and selective time-attention fusion (STA).

As shown in Fig. \ref{fig:f3}, the SCA fusion module consists of max pooling, average pooling, a fully connected (FC) layer, and a sigmoid activation function. Firstly, the input $F$ undergoes max pooling and average pooling along the time dimension to compress the temporal axis, resulting in $F_{max}$, $F_{avg}$, and their element-wise addition feature $F_{add}$. Here, $F_{max} \sim F_{max} \sim F_{add} \in \mathbb{R}^{B*C*1}$. Each feature is then passed through an FC layer followed by a sigmoid activation function.
Finally, each attention representation is weighted and added separately before being activated to obtain the channel attention fused feature $F^{'}$. 

As shown in Fig. \ref{fig:f4}, the STA fusion module consists of max pooling, average pooling, a 1D convolutional layer, and a sigmoid activation function. Firstly, the input F undergoes max pooling and average pooling along the channel dimension to compress the channel axis, resulting in $F_{max}$, $F_{avg}$, and a concatenated feature $F_{concet}$. Here, $F_{max}\sim F_{min} \in \mathbb{R}^{B*1*T}$, and $F_{concet} \in \mathbb{R}^{B*2*T}$. Each feature is then passed through a 1D convolutional layer followed by a sigmoid activation function.
Finally, each attention representation is weighted and added separately before being activated to obtain the time attention fused feature $F^{'}$.

Where $\alpha_{max}$, $\alpha_{avg}$, $\beta$ are hyperparameters empirically set to 0.25, 0.25,  and 0.5, respectively.

\section{EXPERIMENT}
\subsection{Dataset}
We evaluated the performance of speech enhancement using the proposed BSS-CFFMA on the VoiceBank-DEMAND \cite{VB} and WHAMR! \cite{WHAMR} datasets, respectively. The VoiceBank-DEMAND dataset consists of a total of 11572 utterances, with 28 speakers and 824 utterances from 2 speakers used as training and testing sets, respectively. During the training phase, mix 10 types of noise with a signal-to-noise ratio (SNR) of [0, 5, 10, 15] dB with pure speech. During the testing phase, 5 types of noise were mixed with clean speech, with signal-to-noise ratios of [2.5, 7.5, 12.5, 17.5] dB. The WHAMR! dataset is an extended version of the wsj0-2mix \cite{wsj0} dataset, which includes noise and reverberation. Noise is collected from real environments, and the reverberation time is selected to simulate typical home and classroom environments. 
\begin{figure}[htbp]
\begin{minipage}[b]{0.48\linewidth}
  \centering
  \centerline{\includegraphics[width=4.0cm]{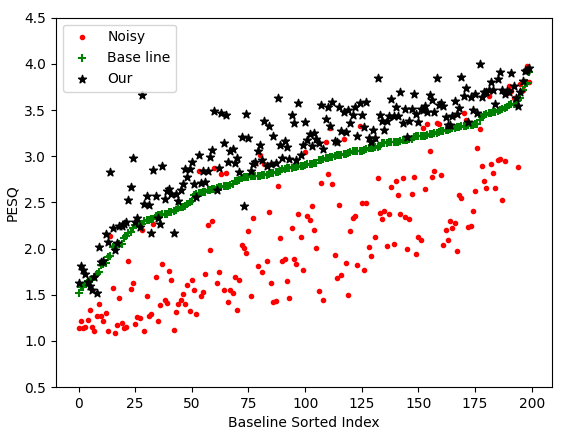}}
%  \vspace{1.5cm}
  \centerline{(a) VoiceBank-DEMAND}\medskip
\end{minipage}
\hfill
\begin{minipage}[b]{0.48\linewidth}
  \centering
  \centerline{\includegraphics[width=4.0cm]{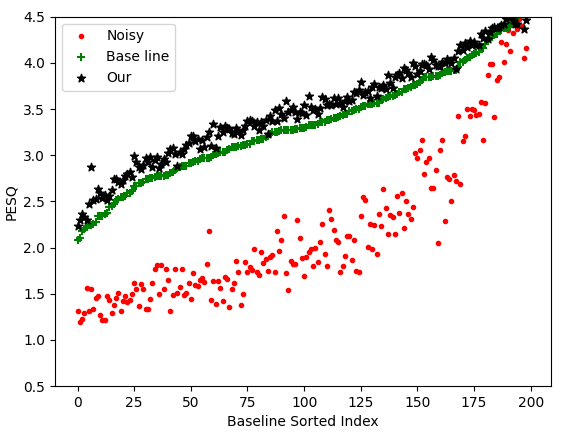}}
%  \vspace{1.5cm}
  \centerline{(b) WHAMR! (Reverb)}\medskip
\end{minipage}
\caption{Pairwise comparison of PESQ with BSS-CFFMA with baseline (BSS-SE) on VoiceBank-DEMAND nad WHAMR! dataset test. Where SSL uses Wav2vec2.0 (without fine-tuning).}
\label{fig:f5}
\end{figure}
Pure speech and noise are randomly mixed within the range of a signal-to-noise ratio of [-6,3] dB. The WHAMR! dataset consists of a training set, a validation set, and a testing set consisting of 20000, 5000, and 3000 voices, respectively.

\subsection{Evaluation Metrics}
In order to evaluate the performance of BSS-CFFMA, we selected the following metrics: wideband perceived assessment of speech quality (WB-PESQ\footnote{PESQ is the same as WB-PESQ \label{s2}}) \cite{PESQ}, narrowband perceived assessment of speech quality (NB-PESQ) \cite{PESQ}, scale-invariant source-to-noise ratio (SI-SNR) \cite{SISNR}, short-time objective intelligibility \cite{STOI}, speech signal distortion prediction (CSIG) \cite{ALL}, background noise invasion prediction (CBAK) \cite{ALL}, overall performance prediction (COVL) \cite{ALL}, and real-time factor (RTF).

\subsection{Experimental Setup}
All speech signals are downsampled to 16 kHz and randomly selected for training 100 rounds with a duration of 2.56 seconds. The STFT and ISTFT parameters are set as follows: FFT length is 25 ms, window length is 25 ms, and hop size is 10 ms. Batch size B is set to 16. We used Adam optimizer and dynamic learning rate strategy \cite{IFFNET}; the learning rate for SSL fine-tuning is 0.1$*$ learning-rate. Train using two Precision T4 GPUs, with a training time average of approximately 7 minutes per epoch.

\section{RESULTS}
\subsection{Performance Comparison on Two Datasets}
In our study, we first compared the denoising performance of the proposed BSS-CFFMA method with 14 baseline methods on the VoiceBank-DEMAND dataset. These methods can be categorized into three different domain approaches. As shown in Table \ref{tab:t1}, the proposed BSS-CFFMA outperforms all the baselines regarding evaluation metrics. In addition, compared to the SSL cross-domain method BSS-SE, BSS-CFFMA significantly surpasses BSS-SE. Even surpassing the performance of BSS-SE on large SSL models in basic SSL models. This result further demonstrates the higher efficiency of our network in leveraging cross-domain features for SSL feature extraction and utilization.
\begin{table}[htbp]
    \centering
    \renewcommand\arraystretch{1.5}
    \caption{Comparison results on the VoiceBank-DEMAND dataset regarding objective speech quality metrics. The SSL models utilize the base Wav2vec2.0 and WavLm (with fine-tuning).}
    \label{tab:t1}
    \resizebox{8.5cm}{!}{
    \begin{tabular}{lcccccc}
        \hline
         \textbf{Methods} & \textbf{Domain} & \textbf{PESQ}$\uparrow$ & \textbf{CSIG}$\uparrow$ & \textbf{CBAK}$\uparrow$ & \textbf{COVL}$\uparrow$ & \textbf{STOI}(\%)$\uparrow$ \\
         \hline
          Noisy  & - & 1.91 & 3.35 &	2.44 &	2.63	& 91.5 \\
          \hline
         SEGAN  \cite{SEGAN}  & Time domain & 2.16 &	3.48 &	2.44 &	2.63	& - \\
         MetricGAN  \cite{METRICGAN}  & T-F domain  & 2.86  & 3.86  & 3.33  & 3.22  & - \\
         WavCRN  \cite{WAVCRN}  & Time domain  & 2.64  &	3.94  &	3.37  &	3.29 & - \\
         MetricGAN+  \cite{METRICGAN+} & T-F domain  & 3.15  & 4.14  & 3.16  & 3.64  & - \\
         CDiffuSE  \cite{CDIFFUSE} & Time domain  & 2.52  & 3.72  & 2.91  & 3.01  & 91.4 \\
         SADNUnet  \cite{SADN-UNet}  & Time domain  & 2.82  & 4.18 & 3.47  & 3.51  & 95.0 \\
         DMF-Net  \cite{DMFNET} & T-F domain  & 2.97 &	4.26 &	3.52	& 3.62 & 94.4 \\
         BSS-SE(wav2vec2.0:Base)  \cite{BSSSE}  & Cross domain  & 2.94 &	4.32 &	3.45 &	3.64 &	94.0 \\
         BSS-SE(WavLM:Base)  \cite{BSSSE} & Cross domain & 3.05 &	4.40	& 3.52 &	3.74 &	95.2 \\
         BSS-SE(WavLM:Large)(PF)  \cite{BSSSE} & Cross domain & 3.20 &	4.53 & 3.60 &	3.88 &	\textbf{95.4} \\
         MANNER(Base)  \cite{MANNER} & Time domain  & 3.12 & 	4.45	&  3.61	 &  3.82 & 	95.0 \\
         FSI-Net  \cite{FSINET}  & T-F domain  & 2.97 &	4.28 &	3.59	& 3.69 & 94.4 \\
         CompNet  \cite{COMPNET}  & T-F domain  & 2.90  & 4.16 & 3.37  & 3.53  & - \\
         SF-Net  \cite{SFNET} & T-F domain  & 3.02 &	4.36 &	3.54	& 3.67 &	94.5 \\
         \hline
        \; BSS-CFFMA(wav2vec2.0:Base) & Cross domain  & 3.09 & 4.43  & 3.61  & 3.80  & 94.2 \\
         \; BSS-CFFMA(wav2vec2.0:Base)(PF) & Cross domain  & 3.15 & 4.46 &	3.66 &	3.84 &	94.5 \\
         \; BSS-CFFMA(WavLM:Base) & Cross domain  & 3.17 & 4.48 & 3.65 &	3.85 &	94.5 \\
        \; BSS-CFFMA(WavLM:Base)(PF) & Cross domain  & \textbf{3.21} & \textbf{4.55} &	\textbf{3.70} &	\textbf{3.91} &	94.8 \\
         \hline
    \end{tabular}
    }
        \begin{threeparttable}
        \begin{tablenotes}
            \footnotesize\tiny
            \item The bold values indicate the best performance for a specific metric.
            \item Large indicates a large number of parameters, while Base indicates a small number of parameters.
            \item PF represents partial fine-tuning.
        \end{tablenotes}
    \end{threeparttable}
\end{table}
\begin{table}[htbp]
    \centering
    \renewcommand\arraystretch{1.8}
    \caption{Comparison results on the WHAMR!  dataset in terms of objective speech quality metrics. The SSL models utilize the base Wav2vec2.0 (with fine-tune).}
    \label{tab:t2}
    \resizebox{8.5cm}{!}{
    \begin{tabular}{l|ccc|ccc|ccc}
        \hline 
        \multicolumn{1}{c} {} & \multicolumn{3}{c}{\textbf{Reverb}} & \multicolumn{3}{c}{\textbf{Noisy}} & \multicolumn{3}{c}{\textbf{Reverb+Noisy}}  \\
         \hline
         \textbf{Methods} & \textbf{PESQ}$\uparrow$  & \textbf{STOI}(\%)$\uparrow$ & \textbf{SI-SNR}$\uparrow$  & \textbf{PESQ}$\uparrow$  & \textbf{STOI}(\%)$\uparrow$ & \textbf{SI-SNR}$\uparrow$ & \textbf{PESQ}$\uparrow$  & \textbf{STOI}(\%)$\uparrow$ & \textbf{SI-SNR}$\uparrow$ \\
         \hline
         Mixed & 2.16 &	91 &	4.38 &	1.11 &	76 &	-0.99 &	1.11 &	73 &	-2.73 \\
         \hline
         PAS-UNet \cite{PASNET} & 3.16 &	- &	\textbf{10.40} &	- &	- &	- &	1.51 &	- &	\textbf{5.33} \\
         DCCRN \cite{IFFNET} & 2.55 &	95 & 7.51 &	1.66 &	90 &	9.03 &	1.59 &	88 & 5.20 \\
         TSTNN \cite{IFFNET} & 2.66 &	95 &	3.56 &	1.94 &	\textbf{93} &	4.17 &	1.91 &	91 &	2.89 \\
         BSS-SE(wav2vec2.0:Base)$^*$ \cite{BSSSE} &3.02  & 91 & 5.90 & 1.84 & 89 & 7.52  &1.70  & 86 &2.16  \\
         \hline
         \; BSS-CFFMA(Wav2Vec2.0:Base) & 3.14 &	95 &	5.97 &	1.92 &	90 & 7.69 &	1.77 & 89 &	2.47 \\
         \; BSS-CFFMA(Wav2Vec2.0:Base)(PF) & \textbf{3.26}	& \textbf{96} & 6.24 &\textbf{2.05} & 92 &\textbf{9.30} & \textbf{1.92} &\textbf{91} &3.55\\
         \hline
    \end{tabular}
    }
    \begin{threeparttable}
        \begin{tablenotes}
            \footnotesize\tiny
            \item * represents the results of the model obtained by our reproduction.
            \item The bold values indicate the best performance for a specific metric. 
            \item PF represents partial fine-tuning
        \end{tablenotes}
    \end{threeparttable}
\end{table}

We also evaluated the denoising, dereverberation, and joint denoising-dereverberation performance of BSS-CFFMA under three test scenarios on the WHAMR! dataset, making it the first self-supervised model used for reverberation tasks. As shown in Table \ref{tab:t2}, BSS-CFFMA outperforms other baselines on most indicators in all three testing scenarios: noise-only (\textbf{Noise}), reverberation-only (\textbf{Reverb}), and simultaneous noise and reverberation interference (\textbf{Reverb+Noise}).

Fig. \ref{fig:f5} provides additional details to aid in the analysis of denoising and dereverberation capabilities across 200 samples. For ease of pairwise comparison, we rank the enhanced speech evaluations according to the baseline BSS-SE model. Through comparison, our model exhibits superior performance in terms of PESQ relative metrics compared to the baseline BSS-SE model.
\begin{table}[htbp]
    \centering
    \renewcommand\arraystretch{1.8}
    \caption{Ablation study on the  VoiceBank-DEMAND dataset. The SSL models utilize the base WavLm (without fine-tune).}
    \label{tab:t3}
    \resizebox{8.5cm}{!}{
    \begin{tabular}{lcccccccc}
        \hline
         \textbf{Methods} & \textbf{WB-PESQ}$\uparrow$ &	\textbf{NB-PESQ}$\uparrow$	& \textbf{CSIG}$\uparrow$ & \textbf{CBAK}$\uparrow$	& \textbf{COVL}$\uparrow$ & \textbf{SI-SNR}$\uparrow$ & \textbf{STOI}(\%)$\uparrow$ &	\textbf{RTF}(Avg)$\downarrow$ \\
         \hline
         Noisy & 1.91 & 2.49 &3.35 &	2.44 &	2.63 & -& 91.5 & - \\
         \hline
        BSS-CFFMA & 3.17 & 3.78 &	4.48 &	3.65 & 3.85 & 19.00& 94.5 & 0.0313 \\
         \hline
       \; w/o MSCFF $\delta$ RHMA (i) &2.90 &3.49 &4.28 &3.43 &3.59 & 17.80 & 93.6 &0.0169  \\
       \; w/o RHMA (ii) &3.08 &3.71 &4.40 &3.60 &3.76 & 18.90 & 94.1 &0.0199   \\
       \; w/o MSCFF $\delta$ MHSA (iii) & 3.07 & 3.70 &4.40	 &3.59	 &3.76	 &18.96  &94.1  & 0.0197 \\
       \; w/o MSCFF $\delta$ SCTA (iV) & 3.10 & 3.72 &4.42 &3.61 &3.79 & 18.94 & 94.2  & 0.0195 \\
       \; w/o MSCFF (v) & 3.14 & 3.76 & 4.46 & 3.62 & 3.84 & 18.68 & 94.4 & 0.0211 \\
         \hline
    \end{tabular}
    }
\end{table}

\begin{figure}[htbp]
\begin{minipage}[b]{0.48\linewidth}
  \centering
  \centerline{\includegraphics[width=4.0cm]{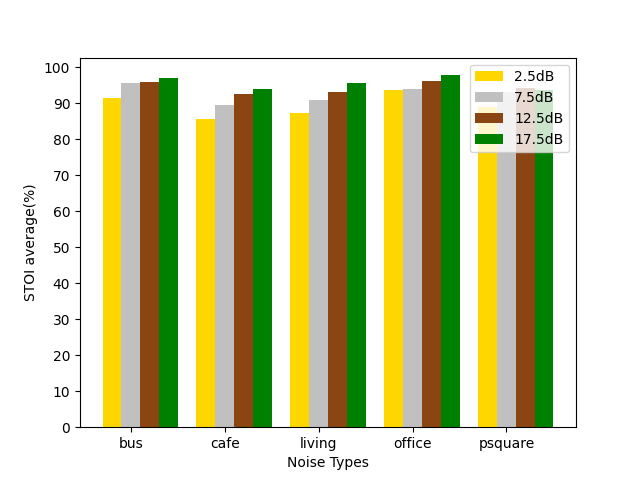}}
  \centerline{(a) STOI score(\%)}\medskip
\end{minipage}
\hfill
\begin{minipage}[b]{0.48\linewidth}
  \centering
  \centerline{\includegraphics[width=4.0cm]{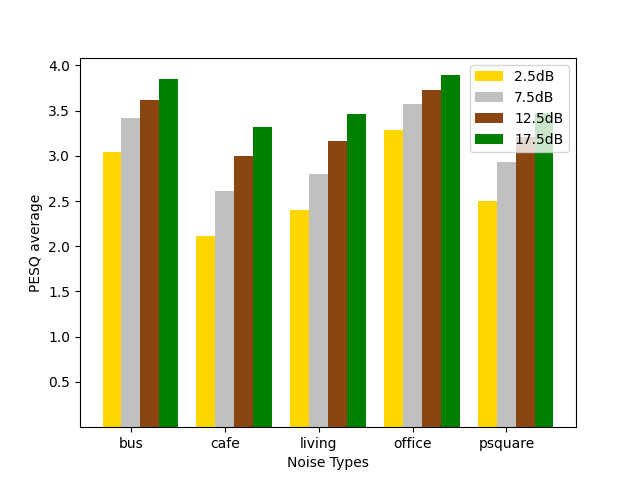}}
  \centerline{(b) PESQ score}\medskip
\end{minipage}
\caption{Comparison in STOI and PESQ average for cases with multiple noise types and different SNR in the VoiceBank-DEMAND dataset. Where SSL models utilize the base WavLM (with fine-tuning).}
\label{fig:f7}
\end{figure}

\subsection{Ablation Analysis}
We conducted ablation experiments to validate each module. Table \ref{tab:t3} shows (i) removing MSCFF and RHMA, (ii) RHMA only, (iii) MSCFF and MHSA, (iv) MSCFF and SCTA, and (v) MSCFF only. All modules outperformed the baseline. PESQ decreased by 0.27 in (i), 0.09 in (ii), 0.10 in (iii), 0.07 in (iv), and 0.03 in (v), demonstrating the effectiveness of MSCFF, RHMA, MHSA, and SCTA.

To further intuitively assess the effectiveness and flexibility of BSS-CFFMA, we conducted additional experiments. Using the test set of the VoiceBank-DEMAND dataset, we categorized the test data according to different noise types and signal-to-noise ratios (SNR) and visualized the PESQ and STOI metrics. Fig. \ref{fig:f7} displays the results of STOI and PESQ across multiple noise types at four SNR levels. We observe that the performance of the network is relatively smooth for different noise-type cases, and there are no extremes in the network for different SNR cases (total average PESQ = 2.7 when SNR = 2.5dB). 
This surface network has relatively good generalization and noise immunity.

\begin{figure}[htbp]
\begin{minipage}[b]{1.0\linewidth}
  \centering
  \centerline{\includegraphics[width=6.0cm]{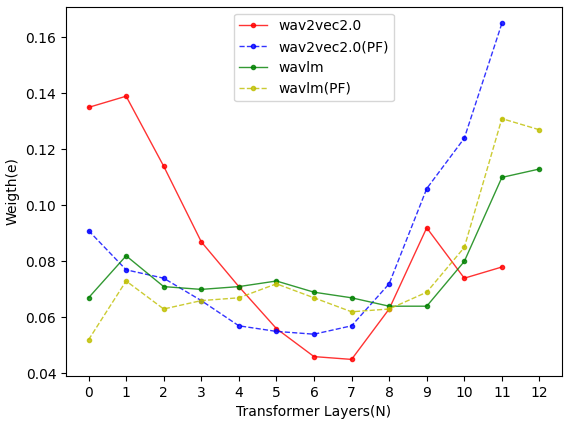}}
\end{minipage}
  \caption{The weighted sum weight corresponding to each layer of SSL Transformer. Wav2vec2.0 has 12 layers (0-11), and WavLm has 13 layers (0-12).}
\label{fig:f6}
\end{figure}

Fig. \ref{fig:f6} presents an analysis of the relationship between the layers of SSL on weighted sum and their corresponding weights. It is observed that irrespective of the SSL model type or fine-tuning status, the weights of the first layer and the last three layers of the SSL model tend to be higher, while the weights of the intermediate layers tend to be lower.

\section{CONCLUSIONS}
In this letter, we propose the BSS-CFFMA model for single-channel speech enhancement and experimentally demonstrate its effectiveness. While it outperforms other baseline models, we also observe that the performance seems to reach a plateau, which we attribute to challenges in phase processing. Therefore, in future work, we will continue to build on our ongoing research and focus on the computation and optimization of phases to improve the model's performance further.

\bibliographystyle{references/IEEEbib}
\bibliography{references/strings,references/refs}

\end{document}